\documentclass[aip,jap,reprint,floatfix,groupedaddress]{revtex4-2}

\usepackage{graphicx}
\usepackage{amsmath,amssymb}
\usepackage{hyperref}

\begin{document}

\title{Transparent electrodes based on mixtures of nanowires and nanorings: A mean-field approach along with computer simulation} 

\author{Yuri~Yu.~Tarasevich}
\email[Corresponding author: ]{tarasevich@asu.edu.ru}
%

\author{Andrei~V.~Eserkepov}
\email{dantealigjery49@gmail.com}

\affiliation{Laboratory of Mathematical Modeling, Astrakhan State University, Astrakhan, 414056, Russia}

\date{\today}

\begin{abstract}
We have studied the electrical conductance of two-dimensional (2D) random percolating networks of zero-width metallic nanowires (a mixture of rings and sticks). We toke into account the nanowire resistance per unit length and the junction (nanowire/nanowire contact) resistance. Using a mean-field approximation (MFA) approach, we derived the total electrical conductance  of these nanowire-based networks as a function of their geometrical and physical parameters. The MFA predictions have been confirmed by our Monte Carlo (MC) numerical simulations. The MC simulations were focused on the case when the circumferences of the rings and the lengths of the wires were equal. In this case, the electrical conductance of the network was found to be  almost insensitive to the relative proportions of the rings and sticks provided that the wire resistance and the junction resistance were equal. When the junction resistance dominated over the wire resistance, a linear dependency of the electrical conductance of the network on the  proportions of the rings and sticks was observed.
\end{abstract}


\maketitle 

\section{Introduction}\label{sec:intro}

Transparent conductive electrodes are an essential component of smart electronics and optoelectronics, such
as smart windows, touch panels and solar cells. Metal nanowire-based transparent conductive electrodes (TCEs), such as Cu, Ag, and Au nanowire electrodes, are considered to be a new generation of TCEs that are capable of replacing indium tin oxide-based electrodes.\cite{Huang2022}

Nanoring-based TCEs are a promising kind of TCE. There exit different methods for synthesizing metallic nanoring-based TCEs. However, in all cases, any TCE includes both nanorings (NRs) and nanowires (NWs). The latter may be bent or wavy. The fraction of NWs and NRs may vary. Table~\ref{tab:NR} presents some published data on synthesized metallic nanorings.
\begin{table*}[!htb]
  \caption{Published  data on metallic nanorings.}\label{tab:NR}
  \centering
  \begin{ruledtabular}
  \begin{tabular}{llll}
Reference & Material & Ring diameter, $\mu$m & Wire diameter, nm \\ \hline
\citet{Zhou2009} & Ag & 4.8--20 & $\approx 165$ \\
\citet{Zhan2011} & Cu & 23 & $\approx 200$ \\
\citet{Yin2015} & Cu & 8--20 & 50 \\
\citet{Azani2018,Azani2019} & Ag  & $15 \pm 5$ & $120 \pm 20$ \\
\citet{Li2020}   & Ag & 7.17--42.94 & 76 \\
\citet{Feng2021} & Ag & 10 & 20--40  \\
\citet{Ning2022} & Ag & 6.54--30.67 & 66.7--115.5 \\
     \end{tabular}
   \end{ruledtabular}
\end{table*}

Notice that, for cylindrical wires of an isotropic metal, the electrical resistivity depends on the wire diameter, tending to the value of the bulk resistance as the wire diameter increases.\cite{Bid2006} Some experimental data are collected in Table~\ref{tab:AgNW}. For comparison, the electrical resistivity of bulk silver is 15.9~n$\Omega\cdot$m.\cite{Serway2013} The temperature dependence of the resistivity should also be borne in mind.\cite{Bid2006,Zhao2020}
\begin{table*}[!htb]
  \caption{Published  data on electrical resistivity of Ag nanowires, ordered in ascending order of nanowire diameter. We used original data on the electrical conductivity\cite{Kojda2015,Wang2018} to calculate the electrical resistivity. 2-P and 4-P are related to two-point probes method and four-point probes method, respectively.}\label{tab:AgNW}
  \centering
  \begin{ruledtabular}
  \begin{tabular}{llllll}
    Reference & Method & $T$, K  & Length, $\mu$m & Diameter $(d)$, nm & $ \rho_\text{w}$, n$\Omega\cdot$m \\
           \hline
\citet{Bid2006} & 2-P & 295 & 6 &  15 & 41.1 \\
\citet{Bid2006} & 2-P & 295 & 6 &  30 & 34.9 \\
\citet{Bellew2015} & 4-P & room temp.  & $7 \pm 2$ & $42 \pm 12 $  & $20.3\pm 0.5$ \\
\citet{Bid2006} & 2-P & 295 & 6 &  50 & 29.2 \\
\citet{Rocha2015} & 4-P & room temp. & 6.7 & $50 \pm 13$ & $22.6 \pm 2.3$ \\
\citet{Zhao2020} & 2-P  &  room temp. & 44 & 84 & 21.3 \\
\citet{Selzer2016} & 4-P & room temp.  & 25 & 90 & $ 31.6 \pm 1.2$\\
\citet{Wang2018} & 4-P & room temp. & 4.88 & 93.2 & 28.5\\
\citet{Wang2018} & 4-P & room temp. & 14.67 & 97.0 & 32.9\\
\citet{Bid2006} & 2-P & 295 & 6 & 100 & 27.8 \\
\citet{Kojda2015} & 4-P & 293 & $15 \pm 1$ & $120 \pm 20$& $37 \pm 13$\\
\citet{Kojda2015} & 4-P & 293 & $11 \pm 1$& $107 \pm 5$ & $29 \pm 5$\\
\citet{Kojda2015} & 4-P & 293 & $13 \pm 1$ & $140 \pm 10$ & $27 \pm 6$ \\
\citet{Kojda2015} & 4-P & 293 & $14 \pm 1$ & $150 \pm 3$ & $25.1 \pm 1.3$\\
\citet{Bid2006} & 2-P & 295 & 6 & 200& 23.3 \\
\citet{Cheng2015} & 2-P & 290 & 27.23 & 227& 79.1 \\
     \end{tabular}
   \end{ruledtabular}
\end{table*}

\citet{Selzer2016} reported that a value of the resistance of a single Ag nanowire (mean diameter of 90~nm) is $4.96 \pm 0.18\,\Omega/\mu$m (the values in the corresponding row in Table~\ref{tab:NR} were calculated using these data) while the junction resistance is $25.2 \pm 1.9\,\Omega$ (annealed junctions) and $529 \pm 239\,\Omega$ (non-annealed ones). For AgNWs (average diameter of $70 \pm 10$~nm and average length of $8 \pm 3~\mu$m), \citet{Charvin2021}, knowing the experimental sample sheet resistance, and expecting the simulated one to be the same, estimated the junction resistance as $14 \pm 2~\Omega$. \citet{Rocha2015}, comparing simulations with experimental data, reported estimates of the junction resistance from 2.28~$\Omega$ to 152~$\Omega$ ($45 \pm 31~\Omega$) (see Supplementary files in Ref.~\citenum{Rocha2015}).

MC simulations of the electrical conductance have been performed for random 2D systems of conductive sticks\cite{Kim2018,Forro2018,Manning2019,Kim2020,Tarasevich2022a,Tarasevich2022b} and rings.\cite{Azani2019,Tarasevich2021,Tarasevich2022} The electrical properties of such systems have been considered theoretically using a percolation theory,\cite{Zezelj2012} a mean-field approximation  (MFA) approach,\cite{Kumar2017,Forro2018,Azani2019,Tarasevich2021,Tarasevich2022,Tarasevich2022a,Tarasevich2022b} network analysis\cite{Kim2021,Daniels2021}, and an effective medium theory (EMT) approach.\cite{OCallaghan2016}

MC simulations of electrical percolation in thin films with conductive disks and sticks have been performed.\cite{Ni2018} The effective conductance of nanocomposites as a function of their relative
concentrations has also been investigated. A synergistic effect has been reported when the disks and sticks combine properly. The widely used junction resistance dominant assumption (JDA) (see, e.g., Refs.~\citenum{Hicks2009,Hicks2018,Manning2019,Kim2020}) has been used, i.e., both disks and sticks were assumed to have no electrical resistance, while a junction between any two conductive fillers was assumed to be a resistor. The resistance of a stick--stick junction was assumed to be five times larger than that of a disk--disk one, since a contact between any two disks is an area, while a contact between any two sticks is a point contact. The dependencies of the electrical conductance with respect to the mass ratio of the disk to stick and to stick length have been plotted.

Recently, the MFA has been successfully applied to pure nanoring- and to pure nanostick-based random, dense 2D systems.\cite{Tarasevich2021,Tarasevich2022,Tarasevich2022a,Tarasevich2022b} However, to the best of our knowledge, the MFA has not previously been applied to a mixture of conductive fillers having different shapes. The goal of the present study was the application of the MFA to 2D systems consisting of randomly deposited conductive NRs and nanosticks.

The rest of the paper is constructed as follows. Section~\ref{sec:methods} presents our computational and analytical methods, viz., Section~\ref{subsec:simul} describes some technical details of our simulation, Section~\ref{subsec:analytics} is devoted to an analytical consideration using a MFA. In Section~\ref{sec:results}, we present our main results and compare the MFA predictions with computer simulations. Section~\ref{sec:concl} summarizes the main results and suggests possible directions for further study.

\section{Methods}\label{sec:methods}

\subsection{Simulation}\label{subsec:simul}

Two kinds of conductive fillers were used in our simulation, viz, rings with a given radius, $r$, and equiprobably orientated zero-width sticks with a given length, $l$.  The electrical resistance per unit length of each filler was $\rho_\text{w}$.

The fillers of both kinds were randomly placed on an insulating substrate. Their centers were independent and identically distributed within a square domain of size $L \times L$. To reduce the finite-size effect, periodic boundary conditions (PBCs) were applied along both mutually perpendicular directions. Let the number density of the sticks be
\begin{equation}\label{eq:ns}
  n_\text{s} =\frac{N_\text{s}}{L^2},
\end{equation}
while the number density of the rings is
 \begin{equation}\label{eq:nr}
n_\text{r} =\frac{N_\text{r}}{L^2}.
\end{equation}
Since the electrical conductance is our primary interest, the total number density of fillers
\begin{equation}\label{eq:n}
n = n_\text{r} + n_\text{s}
\end{equation}
was above the percolation threshold, $n \geqslant n_\text{c}$, in any case.
For each value of the number density, simulations were performed for different proportions of  rings and sticks. When the desired number density of the fillers was reached, the PBCs were removed, allowing us to consider the model as an insulating film of size $ L \times L $ covered by conductive fillers. Then, superconductive busbars were attached to the opposite borders of the domain. A potential difference, $V_0$, was applied to these busbars. The electrical resistance of each contact (junction) between any two fillers was $R_\text{j}$. The electrical resistance of each contact (junction) between a filler and a busbar was $R_\text{b}$. Both kinds of junctions were assumed to be ohmic. Consider a segment of the conductive filler (either a stick or a ring) between the two nearest junctions belonging to it. If a length of this segment is $a$, then its resistivity is $R_\text{s} = \rho_\text{w} a$. Thus, a random resistor network (RRN) exists. This RRN is irregular with different branch resistances. Applying Ohm's law to each branch and Kirchhoff's point rule to each junction, a system of linear equations (SLEs) can be obtained. Although this SLE is huge, its matrix is sparse, therefore, numerical solution of this SLE does not present significant difficulties. We used the EIGEN library\cite{eigenweb} to solve it.

We used domains of a fixed size $L=32 $, while the characteristic sizes of the fillers were $r=1$ and $l = 2 \pi r$. To efficiently determine the percolation threshold (occurrence of a percolation cluster that spans the system in a given direction), the union-find algorithm~\cite{Newman2000,Newman2001} was used. In our simulations, for the two limiting cases, when only one kind of filler is presented, $n_\text{c} = 0.373 \pm 0.004$ for rings, while $n_\text{c} = 5.641 \pm 0.025$ for sticks.
\begin{figure}[!htbp]
  \centering
  \includegraphics[width=\columnwidth]{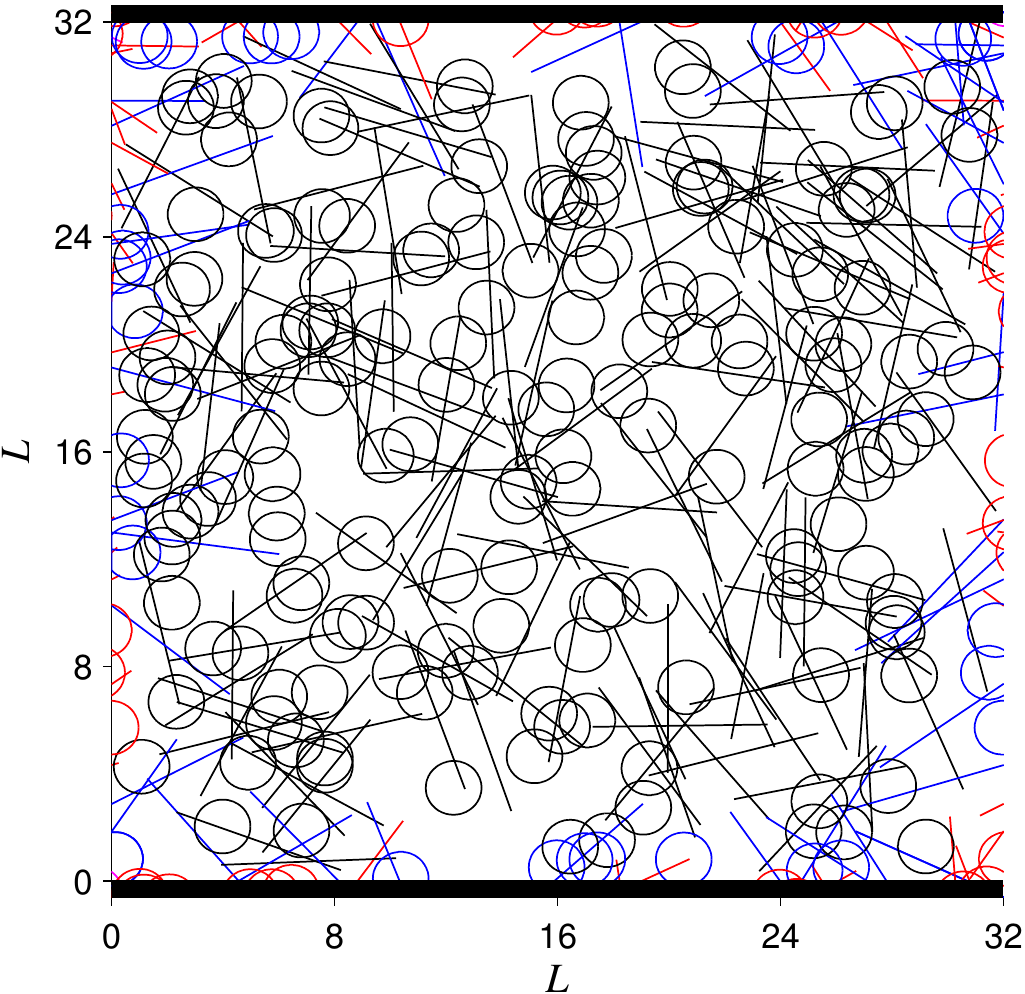}
  \caption{Sample of randomly placed rings and sticks within a domain $L \times L$ with PBCs. All orientations of sticks are equiprobable. The domain size is $L = 32$. The ring radius is $r = 1$. The stick length is $l= 2 \pi r$. The number of rings is $N_\text{r} = 200$. The number of sticks is $N_\text{s} = 200$. Busbars are shown as the thick lines at top and bottom of the system. The effect of the PBCs is demonstrated using color. When a conductive filler intersects a border, its main part (containing the center of the filler) is shown in blue, a part obtained using a translation along only one direction is shown in red, while a part obtained using translations along both directions is shown in magenta.}\label{fig:sample}
\end{figure}

In our study, we concentrated on the experimental data published in Refs.~\citenum{Azani2018,Azani2019}. Tables~\ref{tab:NR} and \ref{tab:AgNW} suggest that in this case $R_\text{w} \approx 125~\Omega$, i.e., of the same order of magnitude or less than the junction resistance, $R_\text{j}$. Therefore, we focused our study on only two cases, viz., JDA and when the resistances of the wires and junctions are equal. Another limiting case when the wire resistance dominates over the junction resistance, considered, e.g., in Refs.~\citenum{Hicks2009,Tarasevich2022b}, is hardly relevant for our system under consideration. In all our computations, we set $R_\text{b} = 0$.

The results of the computations presented in Section~\ref{sec:results} were averaged over 10 independent runs. The error bars in the figures correspond to the standard deviation of the mean. When not shown explicitly, they are of the order of the marker size.

\subsection{Analytical consideration}\label{subsec:analytics}

The probability of the intersection of any two rings is
\begin{equation}\label{eq:Pr}
P_\text{r} = 4 \pi \left(\frac{r}{L}\right)^2
\end{equation}
(see, e.g., Refs.~\citenum{Yi2004,Tarasevich2021}). The probability that a given ring intersects
exactly $N$ other rings is described by the binomial distribution
\begin{equation}\label{eq:ringcontacts}
\Pr(k=N) = \binom{N_\text{r}}{N} P_\text{r}^N(1-P_\text{r})^{N_\text{r} - 1 - N}.
\end{equation}
The expected number of intersections is
\begin{equation}\label{eq:Nrr}
\langle N_\text{rr} \rangle = P_\text{r} (N_\text{r} - 1) \approx 4 \pi r^2  n_\text{r},
\end{equation}
since $N_\text{r} \gg 1$.

The probability of the intersection of any two sticks is
\begin{equation}\label{eq:Ps}
P_\text{s} = \frac{2}{\pi} \left(\frac{l}{L}\right)^2
\end{equation}
(see, e.g., Refs.~\citenum{Yi2004,OCallaghan2016,Kim2018}).
The probability that a given stick intersects exactly $N$ other sticks is described by the binomial distribution
\begin{equation}\label{eq:stickcontacts}
\Pr(k=N) = \binom{N_\text{s}}{N} P_\text{s}^N(1-P_\text{s})^{N_\text{s} - 1 - N}.
\end{equation}
The expected number of intersections is
\begin{equation}\label{eq:Nss}
\langle N_\text{ss} \rangle = P_\text{s} (N_\text{s} - 1) \approx \frac{2}{\pi} l^2  n_\text{s},
\end{equation}
since $N_\text{s} \gg 1$.

\begin{figure}[!htb]
  \centering
  \includegraphics[width=\columnwidth]{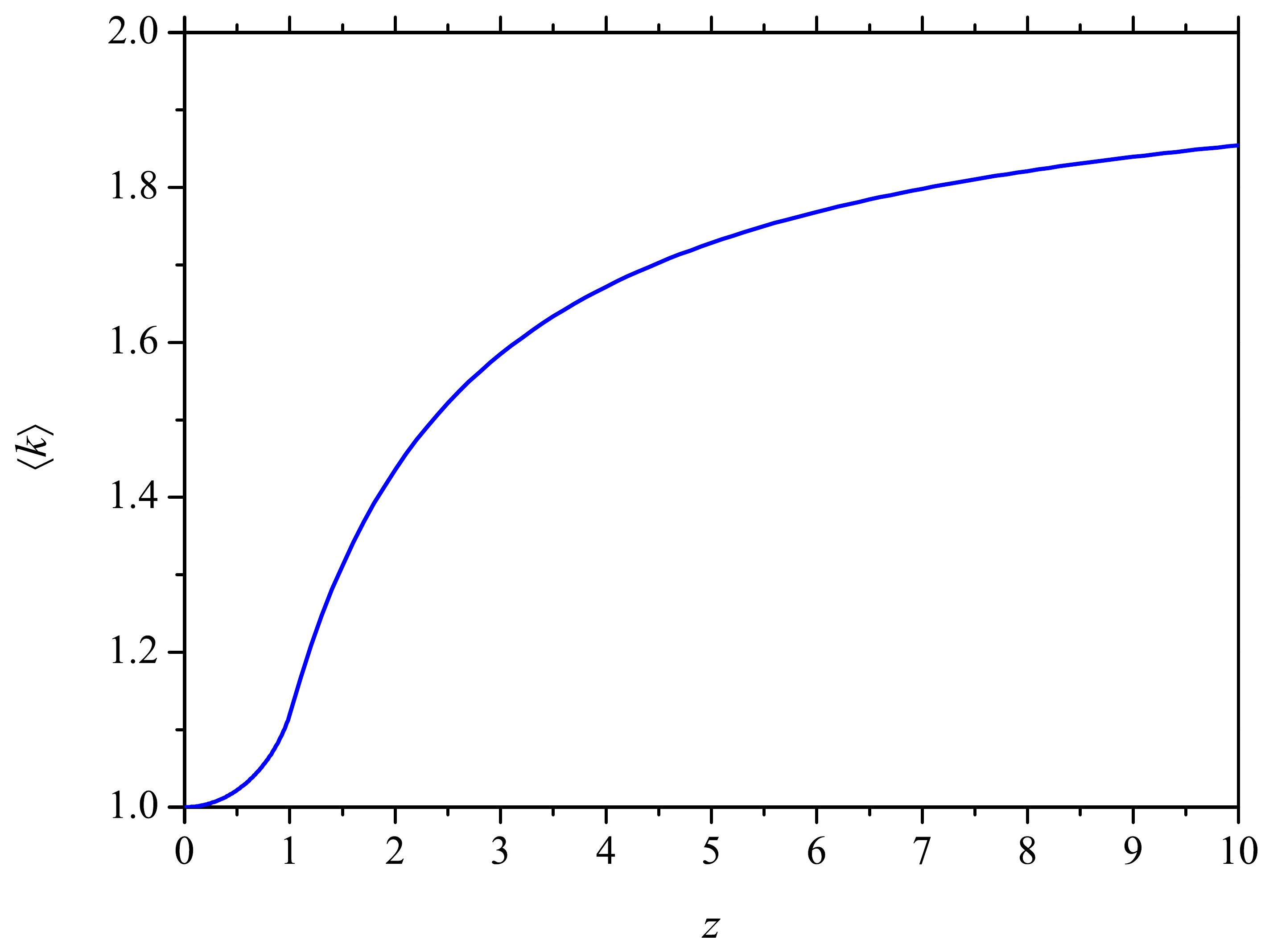}
  \caption{Dependency of the average number of contacts between a stick and a ring, $\langle k \rangle$, on the relative length of the stick, $z$, [Eq.~\eqref{eq:k}].}\label{fig:meank}
\end{figure}

The probability that a stick and a ring have one point of intersection is
\begin{equation}\label{eq:P1}
P_1 =  \frac{ r^2}{ L^2}
  \begin{cases}
  4\left(\arcsin z +z \sqrt{1-z^2}\right), & \text{if } z \leqslant 1, \\
  2\pi, & \text{if } z > 1,
  \end{cases}
\end{equation}
while the probability that a stick and a ring have two points of intersection is
\begin{multline}\label{eq:P2}
  P_2 =\\ \frac{r^2}{L^2}  \left[ 4 z -
\begin{cases}
2 \left(\arcsin z + z \sqrt{1 - z^2}\right), & \text{if } z \leqslant 1, \\
\pi , & \text{if } z > 1,
\end{cases}
\right],
\end{multline}
where
\begin{equation}\label{eq:z}
z = \frac{l}{2r}
\end{equation}
(see Supplement).
The probability of the intersection of a stick and a ring is
\begin{multline}\label{eq:Prs}
P_\text{rs} = P_1 + P_2 =\\
\frac{r^2}{L^2}  \left[ 4 z +
\begin{cases}
 2\left(\arcsin z + z \sqrt{1-z^2} \right), & \text{if } z \leqslant 1, \\
 \pi , & \text{if } z > 1,
\end{cases}
\right].
\end{multline}
Thus, the expected number of intersections of a stick with a ring is
\begin{multline}\label{eq:k}
\langle k \rangle = \frac{P_1 + 2 P_2}{P_\text{rs}} = \frac{ 8 z} { 4 z + \psi}, \quad \text{where }\\
\psi = \begin{cases}
  2 \left(\arcsin z + z \sqrt{1 - z^2}\right), & \text{if } 0 < z \leqslant 1,\\
 \pi, & \text{if }  z > 1 .
\end{cases}
\end{multline}
This quantity varies from $\langle k \rangle = 1$, when $z =0$, through $ 1.6$, when $z=\pi$ (i.e., $l=2\pi r$), to 2, when $z \gg 1$ (Fig.~\ref{fig:meank}).

The probability that a given ring intersects exactly $N$ sticks is described by the binomial distribution
\begin{equation}\label{eq:ringstickcontacts}
\Pr(k=N) = \binom{N_\text{s}}{N} P_\text{rs}^N(1-P_\text{rs})^{N_\text{s} - 1 - N}.
\end{equation}
The expected number of intersections is
\begin{equation}\label{eq:Nrs}
\langle N_\text{rs} \rangle = P_\text{rs} (N_\text{s} - 1) \approx  P_\text{rs} N_\text{s},
\end{equation}
since $N_\text{s} \gg 1$. 

The probability that a given stick intersects exactly $N$ rings is described by the binomial distribution
\begin{equation}\label{eq:stickringcontacts}
\Pr(k=N) = \binom{N_\text{r}}{N} P_\text{rs}^N(1-P_\text{rs})^{N_\text{r} - 1 - N}.
\end{equation}
The expected number of intersections is
\begin{equation}\label{eq:Nsr}
\langle N_\text{sr} \rangle = P_\text{rs} (N_\text{r} - 1) \approx P_\text{rs} N_\text{r} ,
\end{equation}
since $N_\text{r} \gg 1$. 

Thus, the expected number of contacts per ring is
\begin{equation}\label{eq:kr}
\langle k_\text{r} \rangle = 2 \langle N_\text{rr} \rangle + \langle k \rangle \langle N_\text{rs} \rangle,
\end{equation}
while the expected number of contacts per a stick is
\begin{equation}\label{eq:ks}
\langle k_\text{s} \rangle = \langle N_\text{ss} \rangle + \langle k \rangle \langle N_\text{sr} \rangle.
\end{equation}
For any allowed value of $z$,
\begin{equation}\label{eq:kr1}
\langle k_\text{r} \rangle  =
8 \pi r^2  n_\text{r} + 4 l r  n_\text{s},
\end{equation}
\begin{equation}\label{eq:ks2}
\langle k_\text{s} \rangle  =
\frac{2 l^2}{\pi}  n_\text{s} + 4 l r n_\text{r}.
\end{equation}

When the number density of the conductive fillers is large enough, the variation of the electrical potential along the film is close to linear.\cite{Tarasevich2021,Tarasevich2022,Tarasevich2022a,Tarasevich2022b}
Only one conductive filler in the mean-field produced by all the other fillers may be considered, rather than using a consideration of the whole system of fillers. This idea may be easily transferred to the case when the fillers of two different shapes are presented.

Consider a linear conductive wire (stick) in an external electric field. This stick is characterized by a  resistance per unit length, $\rho_\text{w}$. Its lateral surface is supposed to be insulating,  characterized by a leakage conductance per unit length
\begin{equation}\label{eq:Gstick}
G_\text{s} = \frac{R_\text{j} \langle k_\text{s} \rangle}{l}.
\end{equation}
According to Ref.~\citenum{Tarasevich2022a}, the fraction of the electrical conductance, which is due to all sticks, is equal to
\begin{equation}\label{eq:sigmastick}
  \sigma_\text{s} = \frac{ n_\text{s} l }{2  \rho_\text{w}} \left[ 1 - \sqrt{\frac{4 }{ \langle k_\text{s} \rangle \Delta} } \tanh\left(\sqrt{\frac{ \langle k_\text{s} \rangle \Delta}{4} }\right)\right],
\end{equation}
where
\begin{equation}\label{eq:Delta}
  \Delta = \frac{\rho_\text{w} l}{R_\text{j}}.
\end{equation}

Likewise, consider a circular conductive wire (ring) in an external electric field. This ring is characterized by a resistance per unit length, $\rho_\text{w}$. Its lateral surface is supposed to be insulating,  characterized by a leakage conductance per unit length
\begin{equation}\label{eq:Gring}
G_\text{r} = \frac{R_\text{j} \langle k_\text{r} \rangle}{2 \pi r} .
\end{equation}
According to Ref.~\citenum{Tarasevich2021}, the fraction of the electrical conductance, which is due to all rings, is equal to
\begin{equation}\label{eq:sigmaring}
  \sigma_\text{r} = \frac{\pi \lambda^2 r^3 n_\text{r}}{\rho_\text{w} \left( 1 + \lambda^2 r^2 \right)},
  \end{equation}
  where, in our case,
  \begin{equation}\label{eq:lambda2}
  \lambda^2 = \frac{\rho_\text{w}}{R_\text{j}}\left[
4  r  n_\text{r}  + \frac{2 l }{ \pi } n_\text{s}
\right].
  \end{equation}
  The electrical conductance of the system of conductive rings and sticks is
  \begin{equation}\label{eq:sigma}
    \sigma = \sigma_\text{r} + \sigma_\text{s}.
  \end{equation}

When $R_\text{w} \ll R_\text{j}$ (JDA),
\begin{equation}\label{eq:sigmasJDR}
\sigma_\text{s} \approx  \frac{ n_\text{s} l^2 \langle k_\text{s} \rangle}{24 R_\text{j} }.
\end{equation}
For not very large values of the number density, when $\lambda^2 \ll 1$,
\begin{equation}\label{eq:sigmarJDR}
\sigma_\text{r} \approx \frac{\pi r^3 n_\text{r}}{R_\text{j} } \left[
4  r  n_\text{r}  + \frac{2 l }{ \pi } n_\text{s}\right].
\end{equation}
Thus,
\begin{equation}\label{eq:sigmaJDR}
\sigma \approx \frac{1}{R_\text{j} } \left[
4 \pi r^4  n_\text{r}^2  + 2 l r^3 n_\text{r} n_\text{s} + \frac{ n_\text{s}^2 l^4 }{12\pi} + \frac{ n_\text{r} n_\text{s} l^3 r}{6} \right].
\end{equation}
Let $n_\text{s} = x n$, $n_\text{r} = (1-x)n$, then
\begin{multline}\label{eq:sigmaJDRvsx}
\sigma \approx \frac{ n^2 r^4}{R_\text{j} } \times \\
\left[
4 \pi(1-x)^2 +  x (1-x) \left( \frac{2 l}{r} +  \frac{l^3}{6 r^3}\right) +   x^2\frac{l^4 }{12\pi r^4} \right].
\end{multline}
In the particular case, when $l = 2\pi r$,
\begin{equation}\label{eq:sigmaJDRvsxl2pir}
\sigma \approx \sigma_0 \left(1 -  x  + x\frac{\pi^2}{3} \right), \quad \sigma_0 = \frac{4 \pi n^2 r^4}{R_\text{j} }.
\end{equation}
Here, $\sigma_0 $ corresponds to the  electrical conductance of a pure ring system in the case of a JDA~\eqref{eq:sigmarJDR}.

\section{Results}\label{sec:results}
\begin{figure}[!htbp]
  \centering
  \includegraphics[width=\columnwidth]{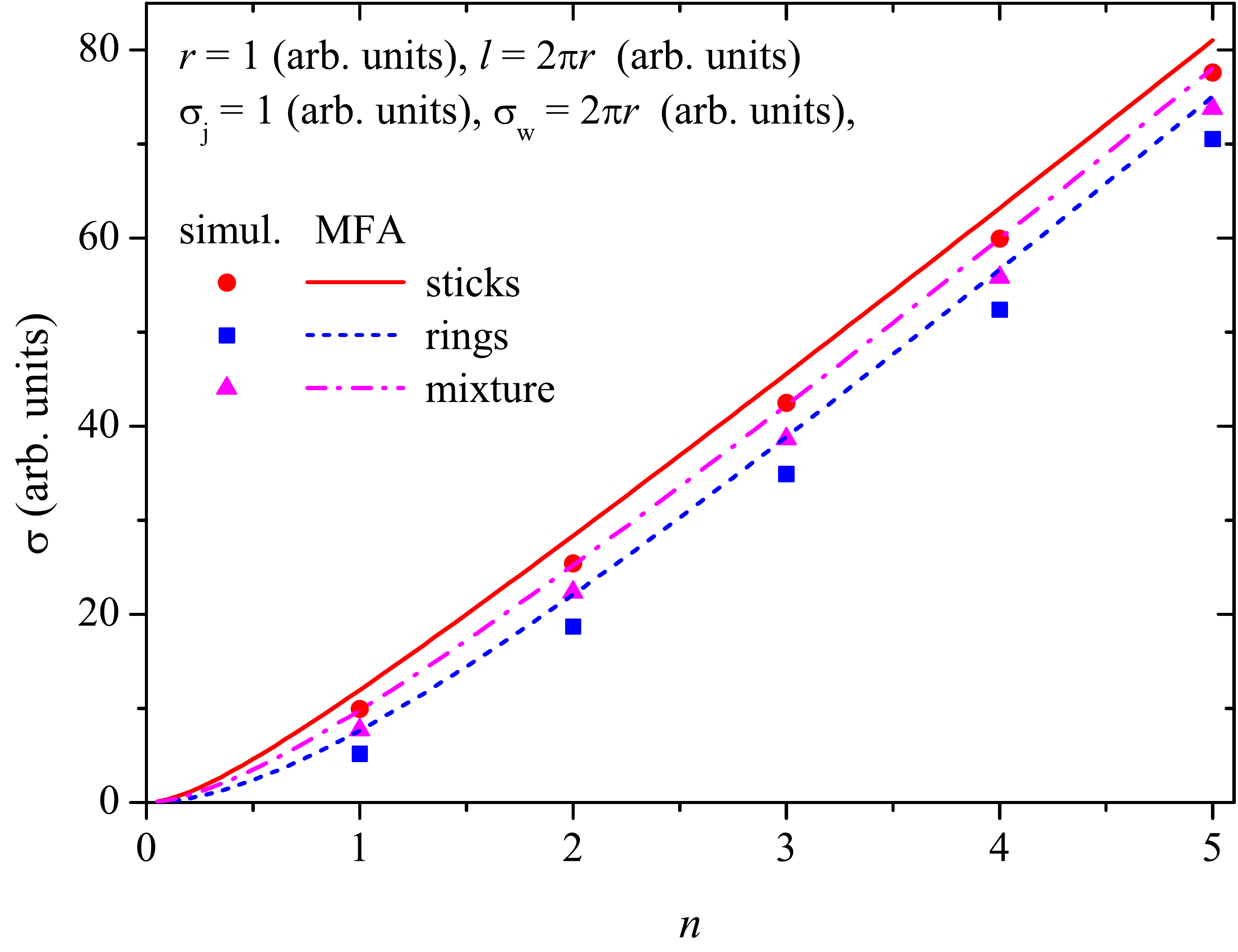}
  \caption{Dependencies of the electrical conductance on the number density of fillers, $n = n_\text{s} + n_\text{r}$, when $n_\text{r} = 0$, $n_\text{s} = n_\text{r}$, and $n_\text{s} = 0$. The MFA predictions are shown as curves, while markers correspond to the simulations. Here, $l = 2\pi r$, $r = 1$, $R_\text{j} = 1$, and $\rho_\text{w} = (2\pi r)^{-1}$.}\label{fig:conductance1}
\end{figure}

Figure~\ref{fig:conductance1} demonstrates the dependencies of the electrical conductance on the number density of the fillers for the three different proportions of rings and sticks (only rings, only sticks, and an equal parts mixture of rings and sticks). The wire resistance and the junction resistance are of the same order. For all used values of the number density, the MFA prediction slightly exceeds the simulated values of the electrical conductance.

\begin{figure}[!htb]
  \centering
  \includegraphics[width=\columnwidth]{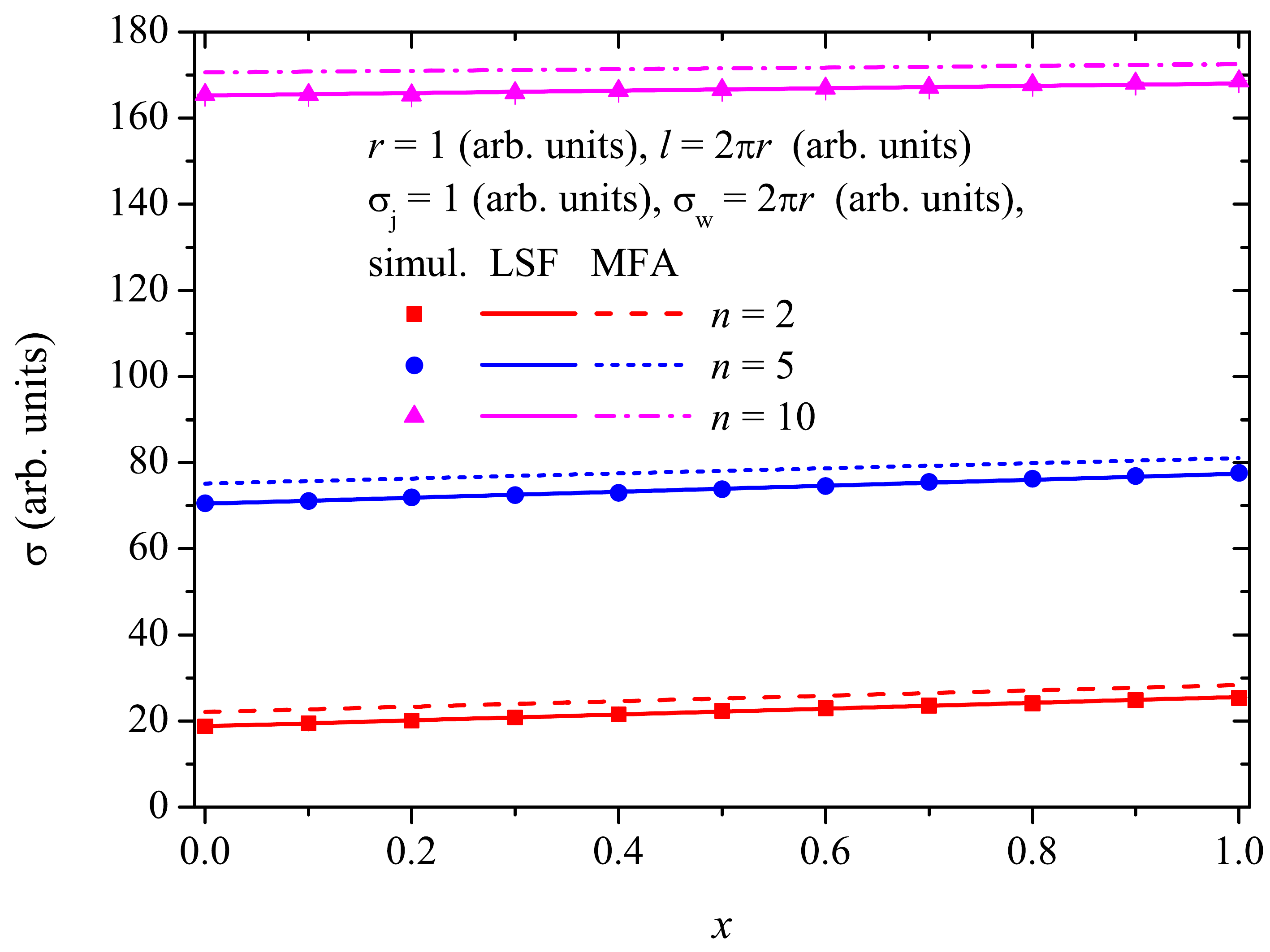}
  \caption{Dependencies of the electrical conductance on the proportions of rings and sticks, $x$, for the three different values of the number density of fillers, $n_\text{s} = x n $ and $ n_\text{r} = (1-x)n$, when $n = 2$, $n = 5$, and $n = 10$.  MFA predictions and the least-squares fitting are shown as lines, while markers correspond to the simulations. Here, $l = 2\pi r$, $r = 1$, $R_\text{j} = 1$, and $\rho_\text{w} = (2\pi r)^{-1}$.}\label{fig:conductance2}
\end{figure}

Figure~\ref{fig:conductance2} presents the dependencies of the electrical conductance on the proportions of rings and sticks, $x$, [$n_\text{r} = x n $ and $ n_\text{s} = (1-x)n$] for the three different values of the number density of fillers, when $n = 2$, $n = 5$, and $n = 10$. The MFA slightly overestimates the electrical conductance, as with the pure stick and pure ring systems.\cite{Tarasevich2021,Tarasevich2022,Tarasevich2022a,Tarasevich2022b} When wire resistance and the junction resistance are of the same order, while $l=2\pi r$, the electrical conductance is almost insensitive to the specific  proportions of rings and sticks.

\begin{figure}[!htb]
  \centering
  \includegraphics[width=\columnwidth]{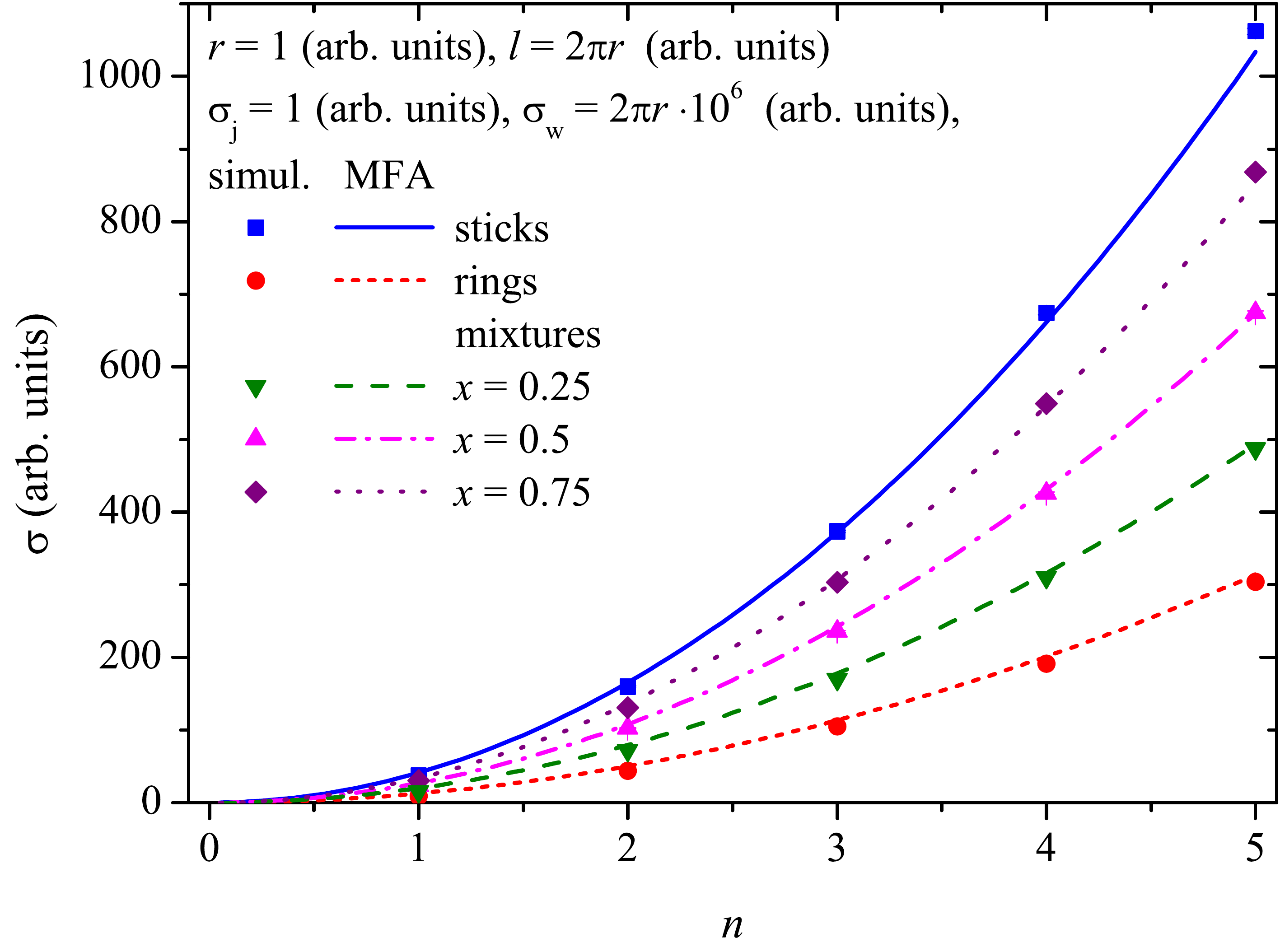}
  \caption{Dependencies of the electrical conductance on the number density of fillers, $n = n_\text{s} + n_\text{r}$, when $n_\text{r} = 0$, $n_\text{s} = n_\text{r}$, $n_\text{s} = 3 n_\text{r}$, $n_\text{r} = 3n_\text{s}$, and $n_\text{s} = 0$. MFA predictions are shown as curves, while markers correspond to the simulations. Here, $l = 2\pi r$, $r = 1$, $R_\text{j} = 1$, and $\rho_\text{w} = (2\pi r)^{-1}10^{-6}$.}\label{fig:conductance3}
\end{figure}

Figure~\ref{fig:conductance3} demonstrates the dependencies of the electrical conductance on the number density of fillers for the five different proportions of rings and sticks. The junction resistance dominates over the wire resistance. MFA predictions and simulations are in good agreement.

\begin{figure}[!htb]
  \centering
  \includegraphics[width=\columnwidth]{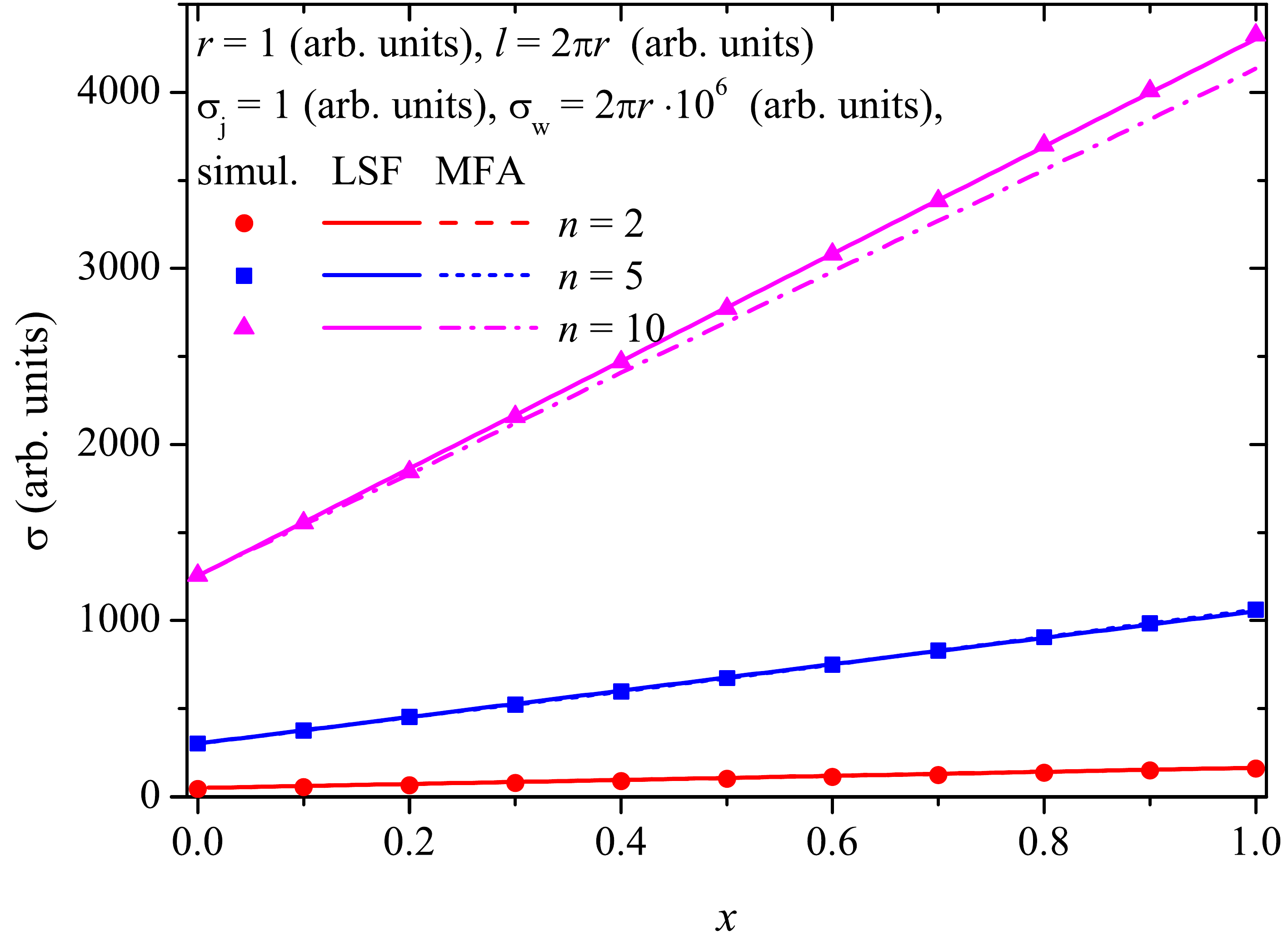}
  \caption{Dependencies of the electrical conductance on the proportions of rings and sticks, $x$, for the three different values of the number density of fillers, $n_\text{s} = x n $ and $ n_\text{r} = (1-x)n$, when $n = 2$, $n = 5$, and $n = 10$.  MFA predictions and the least-squares fitting are shown as lines, while markers correspond to the simulations. For the two larger values of $n$, the difference between the MFA and LSF lines are not visually distinguishable. Here, $l = 2\pi r$, $r = 1$, $R_\text{j} = 1$, and $\rho_\text{w} = (2\pi r)^{-1}10^{-6}$.}\label{fig:conductance4}
\end{figure}

Figure~\ref{fig:conductance4} presents the dependencies of the electrical conductance on the proportions of rings and sticks, $x$, [$n_\text{r} = x n $ and $ n_\text{s} = (1-x)n$] for the three different values of the number density of fillers, viz., $n = 2$, $n = 5$, and $n = 10$, when the junction resistance dominates over the wire resistance. The dependencies are close to linear.

The coefficients of the least-squares fitting of the dependencies of the electrical conductance on the proportions of rings and sticks are collected in Table~\ref{tab:LSF}.
\begin{table}[!htb]
  \caption{Least-squares fitting of the dependencies of the electrical conductance on the proportions of rings and sticks, $\sigma = a x + b$. $l = 2\pi r$, $r = 1$, $R_\text{j} = 1$.}\label{tab:LSF}
  \centering
  \begin{ruledtabular}
  \begin{tabular}{llll}
    $n$ & Slope ($a$) & Intercept ($b$)  & Adj. $R^2$ \\
               \hline
           \multicolumn{4}{c}{$\rho_\text{w} = (2\pi r)^{-1}$}\\
           \hline
           2 & $6.79	\pm 0.09$ & $18.81 \pm 0.05$ & 0.99847 \\
           5 & $6.94 \pm 0.15$ & $70.47 \pm 0.06$ & 0.99542 \\
           10 & $2.76 \pm 0.32$ & $165.3 \pm 0.1$ & 0.88319 \\
                      \hline
           \multicolumn{4}{c}{$\rho_\text{w} = (2\pi r)^{-1}10^{-6}$}\\
           \hline
           2 & 115 & 50 & 1 \\
           5 & $750\pm 3 $ & $302.9 \pm 0.9$ & 0.99981 \\
           10 & $3049 \pm 10$ & $1253.6 \pm 2.5$ & 0.99989\\
           \end{tabular}
    \end{ruledtabular}
\end{table}

\begin{figure}[!htb]
  \centering
  \includegraphics[width=\columnwidth]{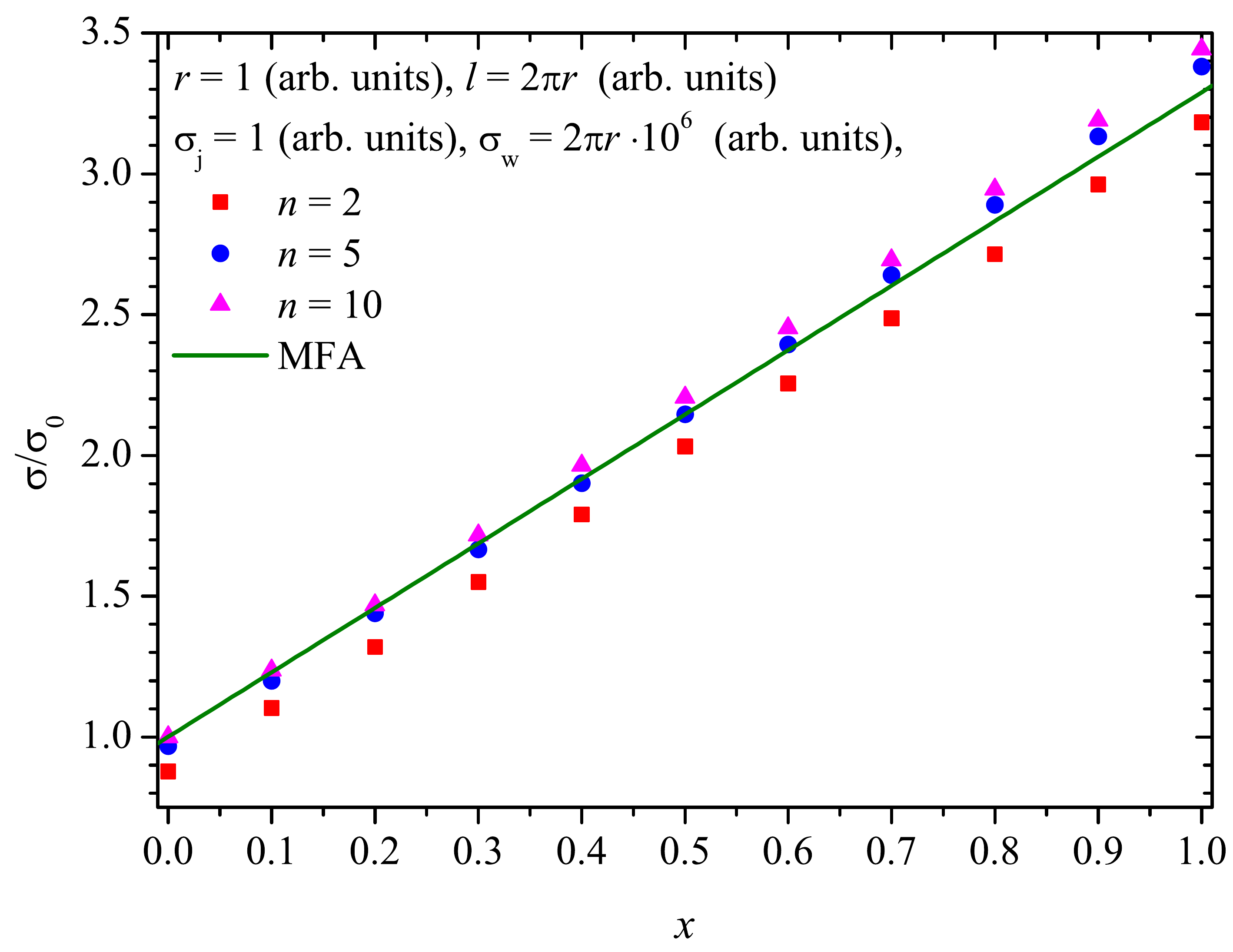}
  \caption{Dependencies of the normalized electrical conductance on the proportions of rings and sticks, $x$, for the three different values of the number density of fillers, $n_\text{s} = x n $ and $ n_\text{r} = (1-x)n$, when $n = 2$, $n = 5$, and $n = 10$. The MFA predictions~\eqref{eq:sigmaJDRvsxl2pir} are shown as a line, while markers correspond to the simulations. Here, $l = 2\pi r$, $r = 1$, $R_\text{j} = 1$, and $\rho_\text{w} = (2\pi r)^{-1}10^{-6}$.}\label{fig:sigmanorm-vs-x}
\end{figure}

Figure~\ref{fig:sigmanorm-vs-x} exhibits the dependencies of the normalized electrical conductance $\sigma/\sigma_0$ on the proportions of rings and sticks, $x$, for the three different values of the number density of fillers in the limiting case of JDA. The linear dependency predicted by an MFA~\eqref{eq:sigmaJDRvsxl2pir} is consistent with the simulations.

\begin{figure}[!htb]
  \centering
  \includegraphics[width=\columnwidth]{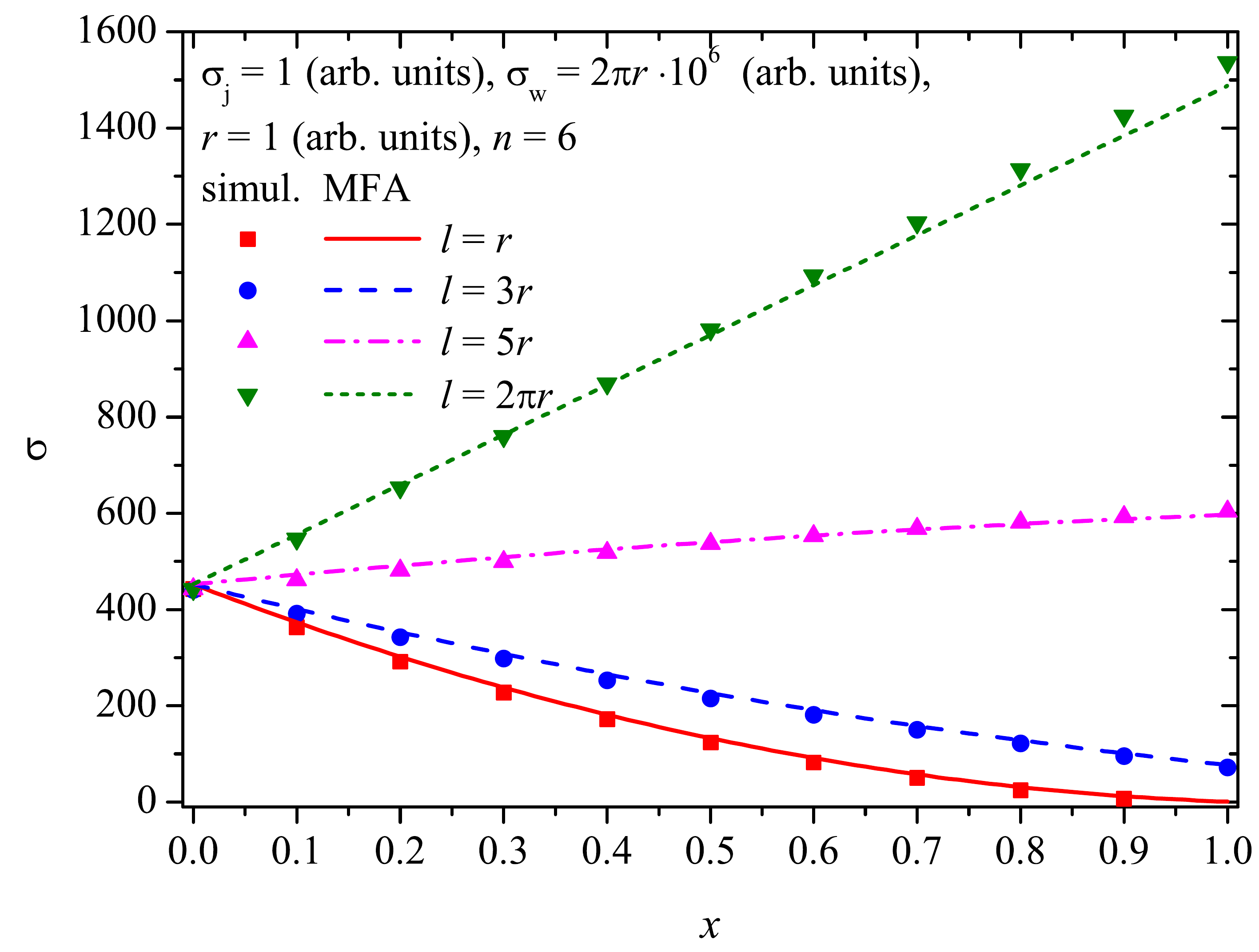}
  \caption{Dependencies of the electrical conductance on the proportions of rings and sticks, $x$, for the fixed values of the number density of fillers and different values of $l$.  MFA predictions are shown as curves, while markers correspond to the simulations. Here,  $r = 1$, $R_\text{j} = 1$, and $\rho_\text{w} = (2\pi r)^{-1}10^{-6}$, $n = 6 $.}\label{fig:Sigma-vs-x-diff-l}
\end{figure}

Figure~\ref{fig:Sigma-vs-x-diff-l} demonstrates the dependencies of the electrical conductance on the proportions of rings and sticks, $x$, for the fixed values of the number density of fillers and different values of $l$, when the junction resistance dominates over the wire resistance (JDA). The MFA predictions~\eqref{eq:sigmaJDRvsx} are shown as curves, while the markers correspond to the simulations.

\begin{figure}[!htb]
  \centering
  \includegraphics[width=\columnwidth]{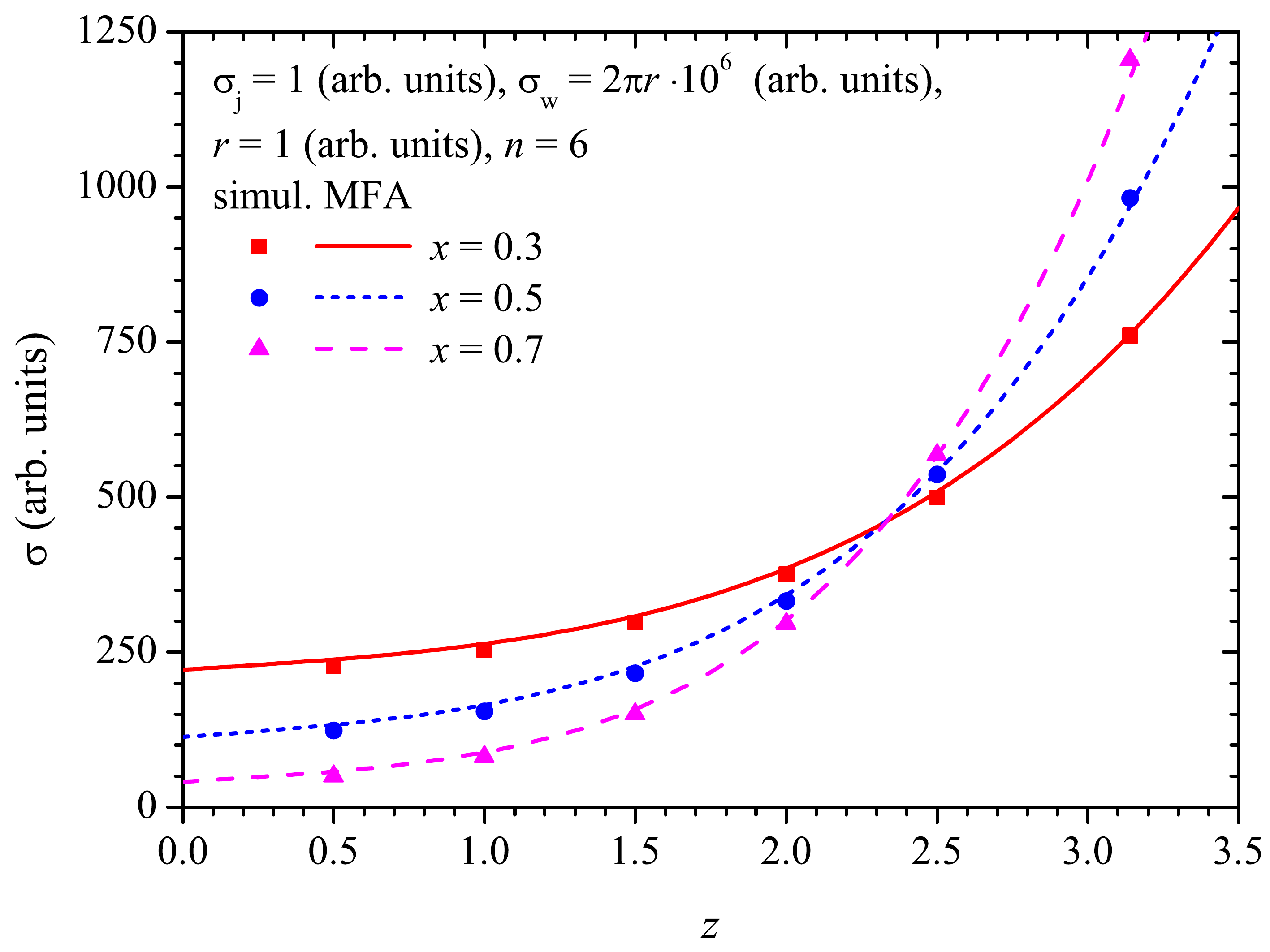}
  \caption{Dependencies of the electrical conductance on the relative stick length, $z$, for the fixed values of the number density of fillers and different values of $x$. The MFA predictions are shown as curves, while the markers correspond to the simulations. Here,  $r = 1$, $R_\text{j} = 1$, and $\rho_\text{w} = (2\pi r)^{-1}10^{-6}$, $n = 6 $.}\label{fig:sigma-vs-z-n6JDR}
\end{figure}

Figure~\ref{fig:sigma-vs-z-n6JDR} demonstrates the dependencies of the electrical conductance on the relative stick length, $z$, for the fixed values of the number density of fillers and different values of $x$, when the junction resistance dominates over the wire resistance (JDA). The MFA predictions~\eqref{eq:sigmaJDRvsx} are shown as curves, while the markers correspond to the simulations.

\section{Conclusion}\label{sec:concl}

The electrical conductance of 2D random percolating networks of zero-width metallic nanowires (a mixture of rings and sticks) has been studied using both a mean-field approximation and a computer simulation. We took into account the nanowire resistance per unit length and the junction (nanowire/nanowire contact) resistance (the so-called multi-nodal representation, MNR~\cite{Rocha2015,Manning2019}). We derived the total electrical conductance of the nanowire-based networks as a function of their geometrical and physical parameters. Our MC simulations were focused on the case when the circumferences of the rings and the lengths of the wires were equal. Our MC simulations confirmed the MFA predictions. The electrical conductance of the network is almost insensitive to the proportions of rings and sticks when the wire resistance and the junction resistance are equal. When the junction resistance dominates over the wire resistance,  a linear dependency of the electrical conductance of the network on the  proportions of rings and sticks was observed. The dispersity of the physical parameters and sizes inevitably  presented in any real-world system can hardly affect the main results of the MFA since only the mean values of all the quantities are important within the MFA. The effect of the dispersity on the results of the simulation is expected to be only an increase in the standard deviation.

There are, at least, two obvious directions of further study. First, since real-world NWs are often bent, more realistic shapes of NWs should be considered instead of sticks. Although the percolative and electrical properties of bent and waved NWs have been studied,\cite{Yi2004,Hicks2018,Langley2018,Esteki2021,Lee2021} there are a number of open questions to be solved. In particular, a mixture of arcs of different curvature up to closed-up arcs (rings) has to the best of our knowledge not yet been studied. Second, direct comparison with experimental data is necessary. Such a comparison requires, simultaneously, data on the wire resistivity, the junction resistance, the transparency, the sheet resistance, the geometrical parameters of both NWs and NRs, and the composition (NR to NW ratios) of the samples. However, direct measurements of the junction resistance are currently limited owing to the small number of teams working on these.

\begin{acknowledgments}
We would like to acknowledge funding from the Foundation for the Advancement of Theoretical Physics and Mathematics ``BASIS'', grant~20-1-1-8-1. We thank I.V.~Vodolazskaya for our fruitful discussions.
\end{acknowledgments}

\section*{Author declarations}
\subsection*{Conflict of Interest}
The authors have no conflicts to disclose.

\subsection*{Author Contributions}
Y.Y.~Tarasevich performed analytical consideration, while A.V.~Eserkepov created the necessary software and performed simulation.
\textbf{Yuri Yu. Tarasevich:} Conceptualization (lead); Formal analysis (equal); Funding acquisition (lead); Methodology (equal); Project administration (lead); Supervision (lead); Visualization (lead);
Writing – original draft (lead); Writing – review \& editing (lead); Investigation (equal).
\textbf{Andrei V. Eserkepov:} Investigation (equal); Software (lead); Validation (equal); Visualization (supporting); Writing – original draft (supporting); Writing – review \& editing (supporting).

\section*{Data availability}
The data that support the findings of this study are available from the corresponding author upon reasonable request.

\bibliography{RingsANDsticks}

\end{document}